\documentclass[11pt]{article}
\usepackage[T1]{fontenc}
\usepackage[utf8]{inputenc}
\usepackage[margin=1in]{geometry}
\usepackage{amsmath,amssymb,graphicx,booktabs,xcolor,microtype}
\usepackage[colorlinks=true,linkcolor=blue,citecolor=blue,urlcolor=blue]{hyperref}

\newcommand{\GuardK}{\ensuremath{48}} 
\newcommand{\GuardSweeps}{\ensuremath{8}} 
\newcommand{\GuardWVel}{\ensuremath{800000}} 
\newcommand{\GuardWAcc}{\ensuremath{3600000}} 
\newcommand{\GuardWPen}{\ensuremath{10000000}} 
\newcommand{\GuardFps}{\ensuremath{30}} 
\newcommand{\GuardZeroStep}{\ensuremath{0}} 
\newcommand{\GuardXi}{\ensuremath{0}} 
\newcommand{\GuardInfluence}{\ensuremath{0.02}} 
\newcommand{\GuardRho}{\ensuremath{0.5}} 
\newcommand{\GuardWHand}{\ensuremath{1000000}} 
\newcommand{\GuardWHandRot}{\ensuremath{50}} 
\newcommand{\GuardWFoot}{\ensuremath{30}} 
\newcommand{\GuardWContact}{\ensuremath{1000000}} 
\newcommand{\GuardSpringHz}{\ensuremath{0}} 
\newcommand{\GuardEps}{\ensuremath{0.002}} 
\newcommand{\GuardFade}{\ensuremath{1}} 
\newcommand{\GuardPullTravel}{\ensuremath{0.06}} 
\newcommand{\GuardRowAlpha}{\ensuremath{1}} 
\newcommand{\GuardWPosVel}{\ensuremath{32000000}} 
\newcommand{\GuardWPosAcc}{\ensuremath{80000000}} 
\newcommand{\GuardWRes}{\ensuremath{100}} 
\newcommand{\GuardSpineCost}{\ensuremath{1}} 
\newcommand{\GuardSpineLimit}{\ensuremath{3}} 
\newcommand{\GuardNeckLimit}{\ensuremath{3}} 
\newcommand{\GuardHipLimit}{\ensuremath{8}} 
\newcommand{\GuardWristLimit}{\ensuremath{8}} 
\newcommand{\GuardArmLimit}{\ensuremath{15}} 
\newcommand{\GuardLegLimit}{\ensuremath{8}} 
\newcommand{\GuardDensity}{\ensuremath{1}} 
\newcommand{\GuardDenseNear}{\ensuremath{0.02}} 
\newcommand{\GuardLimitGain}{\ensuremath{3}} 
\newcommand{\GuardMaxSpeed}{\ensuremath{25}} 
\newcommand{\GuardMaxAcc}{\ensuremath{60}} 
\newcommand{\GuardSlackCap}{\ensuremath{0.01}} 
\newcommand{\GuardHandRelax}{\ensuremath{2}} 
\newcommand{\GuardClearanceBand}{\ensuremath{10}} 
\newcommand{\GuardReleaseTau}{\ensuremath{10}} 
\newcommand{\GuardSnapDeg}{\ensuremath{0.03}} 
\newcommand{\GuardSourceAllow}{\ensuremath{0}} 
\newcommand{\GuardRhoHand}{\ensuremath{0.1}} 
\newcommand{\GuardRhoForearm}{\ensuremath{0.1}} 
\newcommand{\GuardSettle}{\ensuremath{12}} 
\newcommand{\GuardUrgDZero}{\ensuremath{0.008}} 
\newcommand{\GuardUrgDOne}{\ensuremath{0.02}} 
\newcommand{\GuardUrgTau}{\ensuremath{10}} 
\newcommand{\GuardUrgSmooth}{\ensuremath{16}} 
\newcommand{\GuardUrgSpeed}{\ensuremath{3}} 
\newcommand{\GuardUrgAcc}{\ensuremath{6}} 
\newcommand{\GuardUrgSlack}{\ensuremath{10}} 
\newcommand{\GuardSearch}{\ensuremath{0.04}}
\newcommand{\GuardNear}{\ensuremath{0.012}}
\newcommand{\GuardPlaneCos}{\ensuremath{0.7}}
\newcommand{\GuardQueryPurity}{\ensuremath{0.9}}
\newcommand{\GuardColliderPurity}{\ensuremath{0.6}}
\newcommand{\GuardArmQueryStart}{\ensuremath{0.35}}
\newcommand{\GuardSparseFloor}{\ensuremath{8}}
\newcommand{\GuardSparseDivisor}{\ensuremath{4}}
\newcommand{\GuardArmSampleLo}{\ensuremath{0.2}}
\newcommand{\GuardArmSampleHi}{\ensuremath{0.8}}
\newcommand{\GuardArmRatioCap}{\ensuremath{0.9}}
\newcommand{\GuardArmNeutral}{\ensuremath{15}}
\newcommand{\GuardArmCap}{\ensuremath{75}}
\newcommand{\GuardPairAllowance}{\ensuremath{0.003}}
\newcommand{\GuardThighAllowance}{\ensuremath{0.04}}
\newcommand{\GuardContactGap}{\ensuremath{0.008}}
\newcommand{\GuardContactRun}{\ensuremath{12}}
\newcommand{\GuardPullRest}{\ensuremath{0.01}}
\newcommand{\GuardActivationCutoff}{\ensuremath{1\times 10^{-5}}}
\newcommand{\GuardRowTouchCutoff}{\ensuremath{1\times 10^{-9}}}
\newcommand{\GuardInnerSteps}{\ensuremath{8}}
\newcommand{\GuardSkinSearch}{\ensuremath{0.12}}
\newcommand{\GuardSurfaceSupport}{\ensuremath{0.03}}
\newcommand{\GuardContactTemperature}{\ensuremath{0.004}}
\newcommand{\GuardSkinOnset}{\ensuremath{0.002}}
\newcommand{\GuardFacingOnset}{\ensuremath{0.05}}
\newcommand{\GuardDepthGain}{\ensuremath{1.15}}
\newcommand{\GuardWidthDepth}{\ensuremath{0.08}}
\newcommand{\GuardHandWidth}{\ensuremath{0.035}}
\newcommand{\GuardHandWidthGain}{\ensuremath{1.5}}
\newcommand{\GuardHandWidthDepth}{\ensuremath{0.012}}
\newcommand{\GuardBlendScale}{\ensuremath{100}}
\newcommand{\GuardFieldCutoff}{\ensuremath{1\times 10^{-8}}}
\newcommand{\GuardSkinDepthCutoff}{\ensuremath{1\times 10^{-7}}}
\newcommand{\GuardAreaNumerator}{\ensuremath{1.6}}
\newcommand{\GuardAreaIterations}{\ensuremath{16}}
\newcommand{\GuardAreaTolerance}{\ensuremath{0.9999}}
\newcommand{\GuardSkinRim}{\ensuremath{0.035}}
\newcommand{\GuardMaxLookahead}{\ensuremath{5}}
\newcommand{\GuardCandidates}{\ensuremath{4}}
\newcommand{\GuardJointWeights}{\ensuremath{(2,2,8,2,2,8,3,3,8,3,3,8,4,4,8,1,1,1,1,2,2)}}
\newcommand{\GuardPackTable}{%
\begin{tabular}{lrrrr}
\toprule
Part & Query cap & Collider cap & $f_a$ & $D_a$ (m) \\
\midrule
Torso & $0$ & $1000$ & $0.25$ & $0.3$ \\
Head & $0$ & $350$ & $0.03$ & $0.13$ \\
Left upper arm & $60$ & $250$ & $0.35$ & $0.1$ \\
Right upper arm & $60$ & $250$ & $0.35$ & $0.1$ \\
Left forearm & $60$ & $200$ & $0.25$ & $0.08$ \\
Right forearm & $60$ & $200$ & $0.25$ & $0.08$ \\
Left hand & $60$ & $220$ & $0.1$ & $0.05$ \\
Right hand & $60$ & $220$ & $0.1$ & $0.05$ \\
Left thigh & $100$ & $450$ & $0.35$ & $0.18$ \\
Right thigh & $100$ & $450$ & $0.35$ & $0.18$ \\
Left shin & $70$ & $280$ & $0.2$ & $0.1$ \\
Right shin & $70$ & $280$ & $0.2$ & $0.1$ \\
Left foot & $50$ & $220$ & $0.1$ & $0.08$ \\
Right foot & $50$ & $220$ & $0.1$ & $0.08$ \\
\bottomrule
\end{tabular}%
}
\newcommand{\GuardParameterTable}{%
\begin{tabular}{p{.24\linewidth}p{.30\linewidth}p{.35\linewidth}}
\toprule
Parameter & Frozen value & Role \\
\midrule
$K,\; n_{\rm sweep}$ & \GuardK, \GuardSweeps & Rows per endpoint; cyclic sweeps \\
$F$ & \GuardFps & Frames per second \\
$w_r,w_v,w_a$ & \GuardWRes, \GuardWVel, \GuardWAcc & Angular residual and differences \\
$w_{\rm pen},w_{\rm pull}$ & \GuardWPen, \GuardWContact & Hinge weights \\
$w_h,w_o,w_f$ & \GuardWHand, \GuardWHandRot, \GuardWFoot & Wrist position/orientation; feet \\
$w_{pv},w_{pa}$ & \GuardWPosVel, \GuardWPosAcc & Cartesian residual differences \\
$c_{\rm spine}$ & \GuardSpineCost & Spine/neck prior multiplier \\
$\rho,\rho_h,\rho_f$ & \GuardRho, \GuardRhoHand, \GuardRhoForearm & Soft-thickness fractions \\
$I,\epsilon$ (m) & \GuardInfluence, \GuardEps & Activation range; clearance offset \\
$\beta,\alpha$ & \GuardFade, \GuardRowAlpha & Activation and row filter gains \\
$T_p$ (m) & \GuardPullTravel & Pull spatial fade distance \\
$\sigma_{\rm src}$ & \GuardSourceAllow & Additional source-depth allowance \\
$d_{\rm dense}$ (m); mode & \GuardDenseNear, \GuardDensity & Adaptive query density \\
$\Theta_s,\Theta_n$ (deg) & \GuardSpineLimit, \GuardNeckLimit & Spine sum; each neck/head joint \\
$\Theta_h,\Theta_w,\Theta_a,\Theta_l$ (deg) & \GuardHipLimit, \GuardWristLimit, \GuardArmLimit, \GuardLegLimit & Hip, wrist, arm and leg limits \\
$g_\ell,r_h,b_\theta$ & \GuardLimitGain, \GuardHandRelax, \GuardClearanceBand & Shape-dependent clearance controls \\
$V$ (deg/s), $A$ (deg/s$^2$) & \GuardMaxSpeed, \GuardMaxAcc & Residual speed and acceleration \\
$c$ (m) & \GuardSlackCap & Deep-row weight attenuation scale \\
$\tau_r,\theta_{\rm snap}$ (frames, deg) & \GuardReleaseTau, \GuardSnapDeg & Release and exact-zero snap \\
$n_{\rm settle}$ & \GuardSettle & Extra first-call steps \\
$d_0,d_1$ (m) & \GuardUrgDZero, \GuardUrgDOne & Urgency ramp endpoints \\
$\tau_u$ (frames) & \GuardUrgTau & Urgency decay \\
$U_s,U_v,U_a,U_c$ & \GuardUrgSmooth, \GuardUrgSpeed, \GuardUrgAcc, \GuardUrgSlack & Full-urgency schedule factors \\
$z,\xi,\omega_{\rm spring}$ & \GuardZeroStep, \GuardXi, \GuardSpringHz & Disabled shrinkage, approach, spring \\
\bottomrule
\end{tabular}%
}

\newcommand{\NClips}{320}
\newcommand{\NBodies}{3}
\newcommand{\NClipBody}{960}
\newcommand{\NUniformFrames}{19200}

\newcommand{\HandTorsoCopyGtTwenty}{15.53}

\newcommand{\HandTorsoCopyMaxMM}{135.6}
\newcommand{\HandTorsoPoseGtFive}{17.20}

\newcommand{\HandTorsoPoseGtTwenty}{0.18}

\newcommand{\ForearmTorsoCopyGtTwenty}{10.41}

\newcommand{\UpperArmTorsoCopyGtTwenty}{39.79}

\newcommand{\UpperArmTorsoPoseGtTwenty}{37.95}

\newcommand{\NFullFrames}{288{,}000}
\newcommand{\HandTorsoFinalGtFive}{0.05}

\newcommand{\HandTorsoFinalGtTwenty}{0.00}

\newcommand{\HandTorsoFinalMaxMM}{9.3}

\newcommand{\ForearmTorsoFinalGtTwenty}{0.00}

\newcommand{\ForearmTorsoFinalMaxMM}{5.3}

\newcommand{\UpperArmTorsoFinalGtTen}{10.24}
\newcommand{\UpperArmTorsoFinalGtTwenty}{0.11}

\newcommand{\UpperArmTorsoFinalMaxMM}{27.1}
\newcommand{\ScanFinalHandGtTwentyFrames}{97}
\newcommand{\HandOtherCopyGtTwenty}{16.11}
\newcommand{\HandOtherFinalGtTwenty}{0.20}
\newcommand{\ThighCopyGtTwenty}{9.41}
\newcommand{\ThighFinalGtTwenty}{0.17}
\newcommand{\RegressionCount}{0}
\newcommand{\RegressionTotal}{2880}

\newcommand{\ScanConfirmCopyHand}{91.8}
\newcommand{\HFMedian}{1.008}
\newcommand{\HFPNinety}{1.079}
\newcommand{\HFMax}{2.21}
\newcommand{\HandOffsetMeanCM}{0.85}

\newcommand{\UntouchedPct}{81.5}
\newcommand{\ExactFramePct}{25.2}
\newcommand{\ContactKeptCopy}{0.976}
\newcommand{\ContactKeptFinal}{0.953}

\newcommand{\RgenClips}{64}
\newcommand{\RgenFrames}{752}
\newcommand{\LayerOneCausalMeanMS}{0.75}
\newcommand{\LayerOneMeanMS}{1.11}
\newcommand{\LayerOnePNinetyNineMS}{1.66}
\newcommand{\LayerOneMaxMS}{2.16}
\newcommand{\LayerOneOutlierMS}{23.3}
\newcommand{\LayerTwoMeanMS}{1.18}
\newcommand{\LayerTwoPNinetyNineMS}{1.80}

\newcommand{\TotalMeanMS}{2.29}
\newcommand{\LayerTwoVertices}{6890}
\newcommand{\AmassRandomCopyGtTwenty}{3.61}
\newcommand{\AmassRandomPoseGtTwenty}{0.04}
\newcommand{\AmassRandomClips}{120}
\newcommand{\AmassHardCopyGtTwenty}{10.21}
\newcommand{\AmassHardPoseGtTwenty}{0.05}
\newcommand{\AmassHardClips}{80}
\newcommand{\ExtremeCopyGtTwenty}{54.91}
\newcommand{\ExtremePoseGtTwenty}{0.07}
\newcommand{\ExtremeClips}{110}
\newcommand{\ExtremeHandOffsetCM}{5.11}
\newcommand{\ExtremePullKept}{0.449}

\title{Real-Time Self-Penetration Repair for Skinned Characters with Soft Contact Budgets}
\author{Yifei Zhu\\
\href{mailto:zhuyifei.q4@dc.tohoku.ac.jp}{\texttt{zhuyifei.q4@dc.tohoku.ac.jp}}}
\date{}

\begin{document}
\maketitle

\begin{abstract}
Motion that is plausible on one body can penetrate another after a change
in shape, and motion models can produce self-penetration on their native
characters. We present a training-free post-process that reduces these
artifacts through pose correction followed by bounded skin compression.
The module attaches to any motion model whose native character and output
character have compatible character packs, without access to model
internals. Each pack is built offline from the rig and skin without
training. A vertex-level pose layer solves a local quadratic program in
plain C on one CPU core, and a stateless contact deformer acts on the
host's final skinned mesh. A part-dependent soft-tissue budget allocates
correction between pose and skin through a static convention. Source
penetration contributes no extra allowance; optional cloth thickness and
compliance modify the pose layer's collision surface. Operation is causal
or uses a short buffered lookahead, with frame-exact pass-through at the
native interface for clean, inactive frames in the zero state. Across
\NClips{} unseen clips and \NBodies{} body shapes, the fraction of frames
with hand--torso depth above $20\,\mathrm{mm}$ falls from
\HandTorsoCopyGtTwenty\% for copied motion to
\HandTorsoFinalGtTwenty\%, with final maximum depth
\HandTorsoFinalMaxMM\,$\mathrm{mm}$ over every frame. Mean pose and skin
costs are \LayerOneMeanMS{} and \LayerTwoMeanMS\,$\mathrm{ms}$,
respectively, and their sum of \TotalMeanMS\,$\mathrm{ms}$ is a sub-frame
processing cost. The median joint-velocity high-frequency ratio relative
to copied target motion is \HFMedian.
\end{abstract}

\section{Introduction}
\label{sec:intro}

A motion model specifies how a character moves; the rendered skin
determines whether that motion makes geometric sense. Changing torso
width, limb thickness, or body proportions can turn a valid gesture into
a hand inside the chest or an arm crossing the abdomen. Copying joint
rotations preserves their values but changes the spatial relationships
between surfaces. Geometry-aware retargeting addresses this mismatch
\cite{guo2025ultrafast,zhang2023r2et,ye2024meshret}. A second source of
artifacts remains even without a shape change: the model's native motion
can already contain self-penetration. Recent mesh-aware refinement and
post-hoc collision correction explicitly address such inputs
\cite{hong2026artifact,qu2026refm,li2026poseshield}. A reusable repair
module should handle both sources of penetration while preserving the
motion supplied by its host.

The difficulty is to reconcile surface geometry, contact, and temporal
continuity within a streaming budget. Skeletons alone do not describe the
clearance available to a hand beside a thick torso. Conversely, treating
every intersection as a rigid violation can displace a sustained contact
that should instead compress the skin. Contact-aware retargeting has long
recognized that slight intersections can be acceptable in some anatomical
regions \cite{villegas2021contact}. Correction must also release smoothly
after an interaction and recover the exact input when no correction state
remains. A deployment interface should let the host retain its motion
representation, generator, and rendering pipeline.

We introduce a training-free post-process for any motion model whose
native character and output character have compatible character packs.
It treats the model as a black box and operates after generation or
retargeting. An offline builder derives each pack from its rest mesh,
skin weights, skeleton, and semantic joint mapping. At runtime the host
supplies synchronized native and target poses. The native pack remains
necessary even when source penetration has no allowance, because source
observations also identify sustained contacts. No model weights,
gradients, or per-character learning are required.

The module combines two geometric correction scales. Layer~1 updates the
21 non-root body-joint rotations as
$R^{\rm out}_{t,j}=R^{\rm in}_{t,j}
\operatorname{Exp}(\widehat{\delta_{t,j}})$ using a per-frame,
vertex-level QP. Sampled skinned vertices generate local collision rows;
pose tracking and residual temporal costs retain the incoming motion.
Layer~2 then applies bounded, stateless contact displacements to the
host's final skinned mesh. Time-independent skin contact deformation is
established work \cite{brunel2021deformer}; the contribution here is its
integration with pose repair through a static soft-budget convention.

The pose layer discounts a part-dependent soft thickness before charging
penetration. Its default fraction is $\rho=\GuardRho$, with
$\rho_h=\GuardRhoHand$ for hand queries and
$\rho_f=\GuardRhoForearm$ for forearm--torso queries. The source-depth
multiplier is $\sigma_{\rm src}=\GuardSourceAllow$, so a penetrating source
does not authorize additional target penetration. Fixed pair allowances
remain separate. The skin layer re-detects contact and uses the global
$\rho$ to cap compression; it receives no remaining-slack message.
A depth-triggered urgency schedule adjusts smoothing and residual-motion
limits for deeper hand/forearm--torso conflicts. Cloth offsets and a
compliance parameter $\kappa$ modify selected pose collision tests.
These mechanisms operate causally or with short endpoint lookahead;
buffering adds $L/F$ seconds independently of processing time.

Our contributions are:
\begin{itemize}
 \item A model-independent pose-and-skin repair system, with offline,
 training-free native and target character packs, that addresses both
 shape-induced and source-existing penetration.
 \item A static soft-budget convention joining sampled vertex-level pose
 correction to final-mesh contact compression, together with a
 depth-triggered urgency schedule and cloth thickness/compliance.
 \item A single-core implementation with sub-frame processing and
 frame-exact pass-through for clean, inactive zero-state frames at the
 native interface. Existing correction history returns through release;
 geometric clearance alone does not imply an immediate copy.
\end{itemize}

On \NClips{} unseen clips across \NBodies{} body shapes, hand--torso
depth exceeds $20\,\mathrm{mm}$ in \HandTorsoCopyGtTwenty\% of copied
frames and \HandTorsoFinalGtTwenty\% of final frames. The final maximum
depth over every frame is \HandTorsoFinalMaxMM\,$\mathrm{mm}$. Mean pose
and skin costs are \LayerOneMeanMS{} and \LayerTwoMeanMS\,$\mathrm{ms}$,
which sum to \TotalMeanMS\,$\mathrm{ms}$. The median joint-velocity
high-frequency ratio relative to copied target motion is \HFMedian.

\section{Related Work}
\label{sec:related}

\paragraph{Online geometric retargeting.}
Geometric retargeting preserves body-relative relationships as character
proportions change. Molla et al.\ use egocentric surface constraints and
analytic IK for causal, training-free performance animation, with
per-character calibration and capsule geometry
\cite{molla2018egocentric}. Delahaye et al.\ extend this line to finger
contacts using pretrained hand-pose networks
\cite{delahaye2023selfcontact}. Guo et al.\ combine sphere-based semantics
with a fast online solver and explicitly propose post-processing motion
generation \cite{guo2025ultrafast}. Their formulation assumes a
high-quality source and masks implausible source configurations.
ReConForM instead optimizes sparse surface descriptors without training,
processing batches of frames \cite{cheynel2025reconform}. Our module
combines streaming correction with vertex-based observations and treats
penetration already present in the source as repairable input.

\paragraph{Learned geometry and artifact refinement.}
R2ET predicts skeletal and shape residuals; MeshRet transfers dense mesh
interaction fields; STaR models spatial and temporal consistency
\cite{zhang2023r2et,ye2024meshret,yang2025star}. These approaches
integrate geometry into a learned retargeter. Hong et al.\ observe
penetration on the posed mesh and map it through a motion-to-vertex
Jacobian into their learned refinement embedding
\cite{hong2026artifact}. They explicitly demonstrate repair of source
penetration. Their reported batched timings show that learned refinement
can be fast. The described per-frame network suggests causal
compatibility, but streaming is not evaluated. Our distinction is the
training-free, independent interface and measured single-core execution.

\paragraph{Post-hoc motion repair.}
ReFM learns a refinement flow over an initial target motion and accepts
different initialization strategies \cite{qu2026refm}. It needs target
geometry and a trained field, but no source mesh. Its clip-level temporal
attention differs from our causal or endpoint-lookahead operation.
ReFM also reduces penetration relative to references that may themselves
penetrate. Its energy gate can accept a zero update for a whole sequence;
our identity property concerns individual clean, inactive frames in the
zero state. PoseShield provides a generator-agnostic pose projection
using a learned collision constraint, evaluated with fixed body shape;
its motion path additionally optimizes generator input noise
\cite{li2026poseshield}. Thus independent repair and pass-through already
have precedents, with different training, temporal, and interface scopes.

\paragraph{Contact skinning and soft allowance.}
Brunel et al.'s elastic--rigid deformer precedes their time-independent
elastic-contact deformer
\cite{brunel2020elasticrigid,brunel2021deformer}. The latter's abstract
describes independent-frame geometric deformation, local self-contact,
and real-time operation. We use this established skin-stage idea without
claiming equivalence to its detailed algorithm. Galoppo et al.\ already
couple skin contact and skeleton response through simulation
\cite{galoppo2007soft}. Molla and Delahaye's real-time self-contact
pipelines also precede our contact handling
\cite{molla2018egocentric,delahaye2023selfcontact}. Villegas et al.\
distinguish acceptable anatomical intersections and propose paintable
penetration allowances as future work \cite{villegas2021contact}.
Our contribution is the static convention between bounded pose repair
and bounded skin compression, including part-dependent pose allowances
and a cloth collision-surface convention.

\paragraph{QP filters and system scope.}
In robotics, Cai et al.\ use a CBF-QP layer to filter imitation commands
against capsule-based self-collision and human--robot collision
\cite{cai2026safe}. Per-frame QP filtering is therefore an established
structure. Our squared-hinge objective uses sampled skinned geometry
and soft costs; urgency schedules its coefficients and motion bounds
without supplying a barrier certificate. Table~\ref{tab:comparison}
summarizes the closest interfaces. The contribution is their combination
in a training-free pose-and-skin module with native character packs.

\begin{table}[t]
 \centering
 \scriptsize
 \setlength{\tabcolsep}{3pt}
 \begin{tabular}{@{}lcccccccc@{}}
 \toprule
 & \shortstack{Arbitrary\\model}
 & \shortstack{No\\training}
 & \shortstack{Causal/short\\lookahead}
 & \shortstack{Single-core\\real time}
 & \shortstack{Vertex\\level}
 & \shortstack{Source\\repair}
 & \shortstack{Frame-exact\\pass-through}
 & \shortstack{Soft budget\\+ skin} \\
 \midrule
 Guo \cite{guo2025ultrafast} & Stated & Yes & Yes & -- & No & No & -- & No \\
 Hong \cite{hong2026artifact} & No & No & -- & -- & Detection & Yes & -- & No \\
 ReFM \cite{qu2026refm} & Target & No & No & -- & Energy & Partial & -- & No \\
 PoseShield \cite{li2026poseshield} & Pose & No & No & -- & No & Yes & -- & No \\
 ReConForM \cite{cheynel2025reconform} & No & Yes & No & -- & Sparse & -- & -- & No \\
 Ours & Pack & Yes & Yes & Yes & Yes & Yes & State & Yes \\
 \bottomrule
 \end{tabular}
 \caption{Supported properties; -- denotes unestablished evidence.
 Model attachment: Stated is Guo's proposed post-processing use;
 Target is ReFM's target-geometry/trained-field interface; Pose is
 PoseShield's single-pose interface; Pack requires native and target
 character packs. ReFM's vertex energy is used in training, and its
 source repair denotes an aggregate reduction from imperfect references.
 State denotes clean, inactive zero-state frames at the native interface;
 sequence-level zero updates are not frame identity. Our vertex tests
 are sampled in Layer~1 and use the final mesh in Layer~2. The CPU column
 requires evidence of real time on a single core.}
 \label{tab:comparison}
\end{table}

\section{Method}
\label{sec:method}

\begin{figure}[t]
 \centering
 \includegraphics[width=\linewidth]{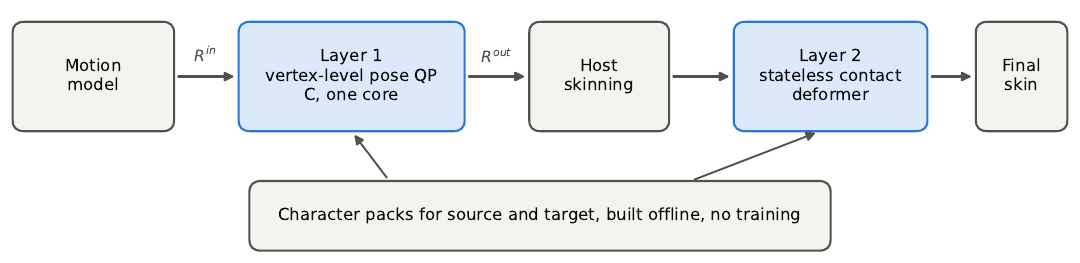}
 \caption{System overview. The host keeps its motion model and skinning.
 The module adds the two blue stages at runtime and one character pack for
 each of the source and target, built offline.}
 \label{fig:pipeline}
\end{figure}

Our plug-in corrects a host-provided humanoid pose after motion generation or
retargeting. Layer~1 maintains a local rotation residual and solves a
box-constrained, squared-hinge quadratic program in plain C, without training.
Layer~2 applies a time-independent contact displacement to the host's skinned
mesh. Time-independent skin contact deformation is an established mechanism
\cite{brunel2021deformer}; here it is coupled to the pose correction through
static soft-thickness budgets. Figure~\ref{fig:pipeline} shows the flow. The
following formulation describes the implemented local surrogates and update
rules.

\subsection{Interface: preserve the host's pose representation}
\label{sec:method-interface}
The host supplies a target character pack, a source character pack, a frame
rate $F$, and synchronized local rotations for both characters. The target
rotations are the upstream model's output; the source rotations describe the
motion on its native character. Both packs use the same ordered semantic body
joints and directed collision pairs, although their skeletons and bind frames
may differ. A source pose can be omitted only when the skeleton hierarchy,
semantic mapping, and local bind bases are compatible with the target.
The interface accepts local rotation matrices or axis--angle vectors.
For each of the 21 non-root semantic body joints,
\begin{equation}
 R^{\rm out}_{t,j}=R^{\rm in}_{t,j}\operatorname{Exp}
 (\widehat{\delta_{t,j}}),\qquad
 \delta_t\in\mathbb R^{63}.
 \label{eq:output}
\end{equation}
Here $\widehat{\cdot}$ is the skew-symmetric matrix map. The root and other
uncontrolled local rotations are copied. Descendants, including fingers,
therefore retain their input local rotations while inheriting corrected
ancestor transforms. Root translation is outside the correction variables;
the pack defines the root offset. Collision geometry uses right-handed,
$Y$-up coordinates in metres and angles in radians.

The host chooses an integer lookahead $0\leq L\leq\GuardMaxLookahead$.
A stream buffers $L$ subsequent inputs before emitting its oldest pose;
Section~\ref{sec:method-lookahead} specifies the endpoint approximation and
startup behavior. After posing the corrected skeleton and skinning its own
mesh, the host may provide final vertices and outward unit normals to
Layer~2, then recompute normals for rendering. A failed native call is an
invalid output, propagated by the interface as an error.

\begin{table}[t]
 \centering\small
 \begin{tabular}{ll}
 \toprule
 Symbol & Meaning \\
 \midrule
 $R^{\rm in},R^{\rm out}$ & Input and corrected local rotations \\
 $d=\delta_{t-1},\ d^-=\delta_{t-2},\ x$ & Previous residuals and current increment \\
 $G_j,p_j,B_j,\ell_j$ & World rotation/position and local bind basis/offset \\
 $c_i,n_i,s_i,\mathsf p(i)$ & Skinned sample, normal, soft thickness, part label \\
 $k=(a,b,\zeta)$ & Directed query/collider part pair and pair class \\
 $g,e,c$ & Raw gap, soft-adjusted gap, final clearance \\
 $A_r,b_r,\omega_r$ & Hinge row, right-hand side, weight \\
 $\mathcal H_t,\ h_t$ & Target endpoints and available lookahead \\
 $u,\ m(u)$ & Urgency state and temporal-cost multiplier \\
 $[z]_+,\ S(z)$ & $\max(0,z)$ and clamped cubic smoothstep \\
 \bottomrule
 \end{tabular}
 \caption{Notation. Joint indices in equations follow the semantic ordering;
 vertex and pair indices refer to their respective character packs.}
 \label{tab:method-notation}
\end{table}

\subsection{Character packs: derive collision geometry from the rig}
\label{sec:method-pack}
Offline preparation takes the rest mesh and outward-wound triangles, skin
weights, a parent-before-child hierarchy, rest joints, and a mapping of the
22 body semantics. Optional world bind bases support rigs whose rest
orientations are not identity. The builder derives local bind bases and
offsets from the world bind frames. Each extra bone inherits the body-part
label of its closest mapped ancestor unless explicit bone labels are given.
Vertices are labeled by the largest accumulated part weight; that weight,
normalized by the total skin weight, is their part purity. The 14 parts are
torso, head, and the left/right upper arms, forearms, hands, thighs, shins,
and feet.

Two independently sampled sets have distinct roles: query vertices test
against collider vertices. Deterministic farthest-point sampling uses purity
thresholds $\GuardQueryPurity$ and $\GuardColliderPurity$, respectively.
Upper-arm queries additionally lie beyond fraction $\GuardArmQueryStart$ of
the shoulder-to-elbow axis. The sample caps and material factors are in
Table~\ref{tab:method-pack}; a part with fewer eligible vertices retains all
of them. The builder retains and renormalizes the four largest skin
influences per vertex, derives area-weighted normals, and estimates soft
thickness from the nearest designated skeletal segment:
\begin{align}
 s_i&=f_{\mathsf p(i)}\min_{(u,v)\in\mathcal S_{\mathsf p(i)}}
 \left\|v_i-\left(q_u+
 \operatorname{clip}_{[0,1]}
 \frac{(v_i-q_u)^\top(q_v-q_u)}{\max(\|q_v-q_u\|^2,\varepsilon_s)}
 (q_v-q_u)\right)\right\|, \\
 G_j(\delta)&=G_{\pi(j)}(\delta)B_jR^{\rm in}_j
 \operatorname{Exp}(\widehat{\delta_j}),\qquad
 p_j(\delta)=p_{\pi(j)}(\delta)+G_{\pi(j)}(\delta)\ell_j, \\
 c_i(\delta)&=\sum_{z=1}^{4}w_{iz}
 \bigl(G_{b_{iz}}(\delta)\,B^{\rm world\,\top}_{b_{iz}}
 (v_i-q_{b_{iz}})+p_{b_{iz}}(\delta)\bigr).
 \label{eq:pack-geometry}
\end{align}
The first line uses rest vertices $v_i$ and semantic rest joints $q_j$;
$\mathcal S_a$ is the fixed segment list for part $a$. A hand uses its wrist
as a degenerate segment. The root uses $G_j=B_jR^{\rm in}_j$ and
$p_j=\ell_j$; uncontrolled joints omit the exponential factor.
Normals use the same blended rotational influences and are normalized.
The small $\varepsilon_s$ is the segment-denominator floor in the builder.

The directed pair list is generated automatically. It includes upper
arm/forearm/hand against torso and head; forearm/hand against thighs, shins,
and the opposite forearm/hand; shin/foot against the opposite leg and torso;
and a directed left-thigh/right-thigh pair. Anatomically adjacent pairs are
not exhaustively enumerated. The full cleaned mesh, topology, labels, and
soft thickness are retained for Layer~2. No per-character collision fit or
training is performed. Isolated invalid vertices/faces are removed and
counted; excessive removal is an error.

\begin{table}[t]
 \centering\small
 \GuardPackTable
 \caption{Automatically sampled pack geometry. $f_a$ scales the estimated
 rest distance to the part skeleton; $D_a$ caps accepted negative planes.
 These constants are extracted from the pack builder.}
 \label{tab:method-pack}
\end{table}

\subsection{Local objective: correct the pose while retaining motion}
\label{sec:method-objective}
Each solve linearizes at the previous residual $d$ applied to the current
input, and optimizes the additive exponential-coordinate increment $x$.
The residual before release is $d+x$; this is not a composition of two
incremental rotation exponentials. The box-constrained problem is
\begin{equation}
 \min_{\underline x\leq x\leq\overline x}
 \frac12x^\top Hx+g^\top x+
 \frac12\sum_r\omega_r[b_r-A_rx]_+^2.
 \label{eq:qp}
\end{equation}
Equivalently, each hinge introduces a nonnegative slack constrained above
$b_r-A_rx$, giving a convex QP. The implementation evaluates the hinges
directly. Collision and contact are soft costs; the component box is hard
within this solve.

Let $\|z\|_W^2=z^\top Wz$. The angular part of the quadratic objective is
\begin{align}
 E_{\rm ang}(x)=\frac12\bigl(&w_r\|d+x\|_W^2
 +m(u)w_v\|x\|_W^2
 +m(u)w_a\|x-(d-d^-)\|_W^2\bigr), \\
 W_{j}&=\nu_jc_j I_3,\qquad
 c_j=\begin{cases}c_{\rm spine},&j\in\{3,6,9,12,15\},\\1,&\text{otherwise,}\end{cases} \\
 (\nu_1,\ldots,\nu_{21})&=\GuardJointWeights.
 \label{eq:angular-cost}
\end{align}
Thus a steady nonzero correction still pays the residual cost. Velocity and
acceleration here are frame differences of the residual, without additional
powers of $F$ in the cost weights. The urgency multiplier $m(u)$ is defined
in Section~\ref{sec:method-urgency}.

For an end effector $e$, define the Cartesian residual
$r_e=p_e(d)-p_e(0)$ and its Jacobian $J_e$ at $d$. The stored
$r_e^-,r_e^{--}$ are the actual residuals at the preceding corrected
outputs, including the change in input pose. Let $\mathcal E_p$ contain
ankles, feet, and wrists, $\mathcal E_h$ the wrists, and $\mathcal E_c$
add shoulders and elbows to $\mathcal E_p$. The remaining quadratic costs
are
\begin{align}
 E_{\rm cart}(x)={}&\frac12\sum_{e\in\mathcal E_p}
 w_e\|r_e+J_ex\|^2 \\
 &+\frac{m(u)}2\sum_{e\in\mathcal E_c}
 \left(w_{pv}\|r_e+J_ex-r_e^-\|^2+
 w_{pa}\|r_e+J_ex-2r_e^-+r_e^{--}\|^2\right), \\
 E_{\rm ori}(x)={}&\frac{w_o}{2}\sum_{e\in\mathcal E_h}
 \|G_e(d)+\mathcal D_e x-G_e(0)\|_F^2.
 \label{eq:cartesian-cost}
\end{align}
Here $w_e=w_f$ for ankles and feet. The shape-dependent wrist weight is
given below. Orientation preservation uses the nine matrix entries,
not a rotation-log distance. $H,g$ are assembled from
$E_{\rm ang}+E_{\rm cart}+E_{\rm ori}$.

Analytic derivatives account for linear blend skinning and additive
exponential coordinates. With $\theta=\|d_j\|$ and
$Q_j=G_j(d)J_r(d_j)$,
\begin{align}
 J_r(d_j)&=I-\frac{1-\cos\theta}{\theta^2}\widehat{d_j}
 +\frac{\theta-\sin\theta}{\theta^3}\widehat{d_j}^{\,2}, \\
 (J_e)_j&=-\widehat{p_e-p_j}\,Q_j\,\mathbf1_{j\preceq e},\qquad
 (\mathcal D_e x)=\sum_{j\preceq e}\widehat{Q_jx_j}\,G_e, \\
 M_{ij}&=\sum_z w_{iz}\mathbf1_{j\preceq b_{iz}}
 (p_{iz}-p_j), \\
 (A_{ab,n})_j&=\bigl((M_{aj}-M_{bj})\times n\bigr)^\top Q_j.
 \label{eq:jacobians}
\end{align}
Here $p_{iz}$ denotes vertex $i$ transformed by influence $z$ before blending.
The relation $j\preceq b$ includes a joint itself. The implementation uses
Taylor expansions near zero. Collision rows freeze the selected normal,
nearest samples, and material budget during each solve; they are local
surrogates rather than derivatives of the complete nearest-surface test.

\subsection{Penetration rows: spend a part-dependent soft budget}
\label{sec:method-penetration}
Transformed influence boxes conservatively bound the sampled skinned
parts. Pair AABB tests determine which query and collider parts are skinned;
exact query bounds prune additional parts. A spatial index then finds a
nearest collider vertex on the requested part. In adaptive density mode,
all queries are used when the pair AABB distance is at most $d_{\rm dense}$.
Otherwise all forearm/hand queries and the first
$\max(\GuardSparseFloor,\lfloor n_a/\GuardSparseDivisor\rfloor)$ queries
of each other part are eligible.

Write $R_s=\GuardSearch$, $r_n=\GuardNear$, and
$\eta_n=\GuardPlaneCos$. For query $a$ and nearest collider $b$, let
$d_{ab}=\|c_a-c_b\|$, $\phi=(c_a-c_b)^\top n_b$, and $o_a$ be the
Euclidean distance from $c_a$ to the collider part AABB. Queries with
$o_a>R_s$ are rejected. The nearest search radius is
$\max(R_s,D_{\mathsf p(b)}/\eta_n)$ if $o_a\leq r_n$, and $R_s$ otherwise.
The signed gap is the following gated normal-plane estimate:
\begin{align}
 \mathcal I_{ab}&=\{\phi<0,\quad
 (d_{ab}<r_n\ \text{or}\ \phi<-\eta_nd_{ab}),\quad
 \phi>-D_{\mathsf p(b)},\quad o_a\leq r_n\}, \\
 g_{ab}&=\begin{cases}\phi,&\mathcal I_{ab},\\d_{ab},&\text{otherwise,}\end{cases}\qquad
 B_{ab}=\rho_{\mathsf p(a),\mathsf p(b)}(s_a+s_b), \\
 \rho_{a,b}&=\begin{cases}
 \rho_h,&a\text{ is a hand},\\
 \rho_f,&a\text{ is a forearm and }b\text{ is torso},\\
 \rho,&\text{otherwise,}
 \end{cases}\qquad
 e_{ab}=\begin{cases}\min(0,g_{ab}+B_{ab}),&g_{ab}<0,\\g_{ab},&g_{ab}\geq0.
 \end{cases}
 \label{eq:effective-gap}
\end{align}
This is a nearest-vertex/normal-plane test, not a mesh signed-distance
function. In particular a rejected negative plane becomes a positive
Euclidean distance. Soft thickness cancels negative gap up to the budget;
it does not enlarge a positive gap.

For pair $k$, the source observation $e^s_{t,k}$ is the minimum effective
gap on the source pack. The additional allowance and candidate clearance are
\begin{align}
 a_{t,k}&=a_k^0+\sigma_{\rm src}[-e^s_{t,k}]_+,\qquad
 a_k^0=\begin{cases}\GuardThighAllowance,&\zeta_k=6,\\
 \GuardPairAllowance,&\text{otherwise,}\end{cases} \\
 c_{t,k,v}&=e_{t,k,v}+a_{t,k}-\epsilon,
 \qquad \sigma_{\rm src}=\GuardSourceAllow.
 \label{eq:allowance}
\end{align}
The frozen source-depth multiplier is zero: source penetration supplies
no additional permission to penetrate. The fixed pair allowance and the
soft budget remain present. Pair minima and candidate lists start at
$R_s$, and only strictly smaller effective gaps replace them.

Each directed pair retains up to $\GuardCandidates$ candidates sorted by
effective gap. Under the per-endpoint budget $K$, the implementation first
selects worst candidates from different pairs in increasing clearance,
then fills remaining slots globally from the other candidates. Only
candidates with $c<I$ enter these two passes. Remaining capacity may retain
previously active row slots while they fade. If $K$ is smaller than the
number of eligible pairs, only the most severe worst candidates fit.

For a selected valid row let
$S(z)=\bar z^2(3-2\bar z)$ with
$\bar z=\operatorname{clip}_{[0,1]}z$. Activation and the absolute-coordinate
plane filter are
\begin{align}
 q_r&=S((I-c_r)/I),\qquad
 \gamma_r\leftarrow\gamma_r+\beta(q_r-\gamma_r), \\
 \widetilde A_r&\leftarrow(1-\alpha_r)\widetilde A_r+\alpha_r A_r,
 \qquad
 \widetilde B_r\leftarrow(1-\alpha_r)\widetilde B_r
 +\alpha_r(-c_r+A_rd), \\
 A_r^{\rm QP}&=\widetilde A_r,\quad
 b_r=\widetilde B_r-\widetilde A_rd,\quad
 \omega_r=w_{\rm pen}\gamma_r
 \min\left(1,\frac{c_{\rm cap}}{\max(c_{\rm cap},-c_r)}\right).
 \label{eq:penetration-row}
\end{align}
Unselected or invalid slots have desired activation zero. A fresh slot
uses $\alpha_r=1$; otherwise $\alpha_r=\alpha$. A retiring slot with no
current candidate retains its old plane. Activations below
$\GuardActivationCutoff$ are removed when their desired value is zero.
The frozen $\beta=\GuardFade$ and $\alpha=\GuardRowAlpha$ make valid
row updates immediate. $c_{\rm cap}$ is the urgency-scheduled attenuation
scale, specified below; it reduces deep-row weights, without imposing a
hard upper bound on slack or on the final hinge gradient. The optional
approach-rate term is disabled by $\xi=\GuardXi$.

\subsection{Contact pull: retain sustained source contacts}
\label{sec:method-contact}
Source contacts produce a separate one-sided closing cost. For every pair
whose query part is a hand, a run counter increments while the source
effective pair gap is below $d_p=\GuardContactGap$ and resets otherwise.
Pull becomes eligible after $n_p=\GuardContactRun$ consecutive observations.
For the target pair's minimum-effective-gap samples, let $\ell$ be their
current distance and $\ell^0$ their distance under the uncorrected target
pose. With rest distance $r_p=\GuardPullRest$,
\begin{align}
 q_k^p&=1-S\left(\frac{[\ell^0-r_p]_+}{T_p}\right),\qquad
 \gamma_k^p\leftarrow\gamma_k^p+\beta(q_k^p-\gamma_k^p), \\
 n_k^p&=-\frac{c_a-c_b}{\ell},\qquad
 A_k^p=A_{ab,n_k^p},\quad b_k^p=\ell-r_p,\quad
 \omega_k^p=w_{\rm pull}\gamma_k^p q_k^p.
 \label{eq:pull}
\end{align}
The gate is zero unless source pull is eligible, a valid candidate exists,
and pull is enabled. Rows are skipped when the current distance vanishes
or the activation is negligible. These are squared hinges in
Eq.~\eqref{eq:qp}: they penalize insufficient closing, rather than enforcing
an equality contact. Pull rows are added after penetration selection and
do not consume $K$. With the frozen fade gain, the weight contains the
square of the spatial gate.

\subsection{Component boxes: bound residual motion and leave room to brake}
\label{sec:method-bounds}
Rest geometry automatically adjusts arm clearance. For each side, the
builder in C measures the outward torso extent $e_s$ from the shoulder
along the inter-shoulder lateral axis, and upper-arm radius $r_s$ over the
axial interval $[\GuardArmSampleLo,\GuardArmSampleHi]$ of arm length
$\ell_s$. Its additional clearance angle and wrist position weight are
\begin{align}
 \chi_s&=\left[\frac{180}{\pi}
 \arcsin\min\left(\GuardArmRatioCap,\frac{e_s+r_s}{\ell_s}\right)
 -\theta_0\right]_+,\qquad \theta_0=\GuardArmNeutral, \\
 \Theta_{a,s}&=\min\bigl(\GuardArmCap,
 \Theta_a+g_\ell[\chi_s-b_\theta]_+\bigr),\qquad
 w_{h,s}=\frac{w_h}{1+r_h[\chi_s-b_\theta]_+/\theta_0}.
 \label{eq:shape-clearance}
\end{align}
Angles in this equation are in degrees. Each of the three spine joints
has limit $\Theta_s/3$; neck and head each use $\Theta_n$, hips use
$\Theta_h$, wrists use $\Theta_w$, remaining arm/collar joints use
$\Theta_{a,s}$, and remaining leg joints use $\Theta_l$.
The inscribed component box uses
$L_j=(\pi/180)\Theta_j/\sqrt3$, so its residual norm is bounded by the
corresponding angular limit. These are bounds on correction, not on the
input pose's absolute anatomical angles.

Let $v_i$ be the stored previous residual increment. The scheduled scalar
component speed $V_t$ and acceleration $A_t$ are converted from degrees per
second and degrees per second squared to radians per frame and per frame
squared. For distance $D\geq0$, the inverse discrete stopping-distance
function is
\begin{align}
 n(D,A_t)&=\left\lfloor\frac{\sqrt{1+8D/A_t}-1}{2}\right\rfloor,
 &\mathcal V(D,A_t)&=\frac{D}{n(D,A_t)+1}+\frac{A_tn(D,A_t)}2, \\
 h_i^-&=-L_j-d_i,&h_i^+&=L_j-d_i, \\
 v_i^-&=\max\{h_i^-,-\min(V_t,\mathcal V(-h_i^-,A_t))\},&
 v_i^+&=\min\{h_i^+,\min(V_t,\mathcal V(h_i^+,A_t))\}, \\
 \underline x_i&=\max\{v_i^-,v_i-A_t\},&
 \overline x_i&=\min\{v_i^+,v_i+A_t\}.
 \label{eq:bounds}
\end{align}
This braking calculation includes the next discrete frame, using
$D=u+[u-A_t]_++[u-2A_t]_++\cdots$.
If urgency decay makes the interval inconsistent, the implementation
instead fixes that component to
$\operatorname{clip}_{[h_i^-,h_i^+]}(v_i-A_t)$ for $v_i>0$, and to
$\operatorname{clip}_{[h_i^-,h_i^+]}(v_i+A_t)$ otherwise. This recovery
preserves the hard residual box while braking; it can temporarily exceed
the newly tightened speed or stopping envelope. Roundoff-size interval
inversions are collapsed to their midpoint.

\subsection{Coordinate descent: solve the bounded hinge surrogate}
\label{sec:method-solver}
The native solver performs $\GuardSweeps$ cyclic sweeps
\cite{wright2015coordinate}, starting from zero projected into the box.
For coordinate $i$, write $r_r=b_r-A_rx$ and $h_i=(Hx+g)_i$.
The one-dimensional derivative and curvature at displacement $z$ are
\begin{align}
 f_i'(z)&=h_i+H_{ii}z-
 \sum_{r:r_r-A_{ri}z>0}\omega_r A_{ri}(r_r-A_{ri}z), \\
 f_i''(z)&=H_{ii}+\sum_{r:r_r-A_{ri}z>0}\omega_r A_{ri}^2.
 \label{eq:coordinate}
\end{align}
At most $\GuardInnerSteps$ safeguarded Newton updates bracket the scalar
minimum inside $[\underline x_i-x_i,\overline x_i-x_i]$.
A proposed coordinate change is rejected if its evaluated objective
increase exceeds the numerical tolerance. Accepted changes update $Hx+g$
and every hinge residual incrementally. All solver iterates satisfy the
box; the fixed sweep count does not assert convergence to the exact QP
minimum. Subsequent release and snapping are separate update rules.

\subsection{Release: reach exact zero and copy untouched input}
\label{sec:method-release}
To remove corrections after contact, the implementation forms a bounded
release proposal and marks joints whose release would worsen a row beyond
its hinge boundary:
\begin{align}
 z_i&=\operatorname{clip}_{[\underline x_i,\overline x_i]}
 (-d_i/\tau_r), \\
 \mathcal J_{\rm held}&=\{j:\exists r,\ A_rz<b_r\ \text{and}\
 \|(A_r)_j\|_\infty>\GuardRowTouchCutoff\},\qquad
 x'_j=\begin{cases}x_j^{\rm QP},&j\in\mathcal J_{\rm held},\\z_j,&\text{otherwise,}\end{cases} \\
 \delta_{t,j}&=\begin{cases}
 0,&j\notin\mathcal J_{\rm held}\ \text{and}\
 \|d_j+x'_j\|<(\pi/180)\theta_{\rm snap},\\
 d_j+x'_j,&\text{otherwise.}
 \end{cases}
 \label{eq:release}
\end{align}
The frozen radial shrinkage parameter is $\GuardZeroStep$, so only this
release/snap rule changes the solved increment. Every assembled row,
including pull and lookahead rows, participates in the held-joint test.
The release proposal respects the computed box; snapping occurs afterward
and does not recheck that box or the rows. This is a deterministic return
to zero, not an additional QP solve or a row-feasibility guarantee.

When a joint residual is exactly zero, the native output copies that
joint's input matrix with \texttt{memcpy}, bypassing exponential evaluation
and matrix multiplication. With zero residual and zero stored residual
velocity, a frame whose current and lookahead geometry both have no nearby
pairs takes an immediate whole-pose copy path. More generally, a zero-state
frame with no active hinge violation also retains zero. Thus clean,
inactive frames pass through bit-identically at the C double-matrix
interface. The axis--angle wrapper restores each untouched joint directly
from the original input values, in the returned double representation.
Geometric clearance alone does not discard a nonzero correction history;
such a joint returns through the release rule.

\subsection{Lookahead and settling: anticipate contact with one residual}
\label{sec:method-lookahead}
For an available horizon $h_t$, target collision and pull rows are assembled
at
\begin{equation}
 \mathcal H_t=\begin{cases}\{0\},&h_t=0,\\\{0,h_t\},&h_t>0,\end{cases}
 \qquad h_t=\min(L,\text{number of queued future frames}).
 \label{eq:horizon}
\end{equation}
Both endpoint geometries apply the same previous residual $d$ and share
the same decision increment $x$. Only the current target pose contributes
the quadratic tracking costs. Intermediate target frames are not
collision-constrained. Source observations for every frame through the
endpoint are cached chronologically, and separate current/future row-slot
states are maintained. This is a single-residual endpoint surrogate,
not a sequence optimization with independent future residuals.

In steady streaming the buffering delay is $L/F$ seconds. Flushing reduces
$h_t$ as the queue empties and introduces no extrapolated poses. On the
first call, the implementation runs $n_{\rm settle}=\GuardSettle$ extra
internal steps on the same supplied window, then one final step whose pose
is emitted. Each internal step advances residual history and the internal
frame counter. After the first internal step, source caching consequently
re-observes the supplied endpoint at each new internal index. Startup
settling therefore advances source contact counters as well as pose
state; it is not an optimization with source time frozen.

\subsection{Depth-triggered urgency: schedule the same QP}
\label{sec:method-urgency}
Urgency allows a deeper hand/forearm--torso conflict to weaken temporal
smoothing and enlarge the residual-motion envelope. It changes coefficients
and bounds of Eq.~\eqref{eq:qp}; it introduces no new collision constraint.
For endpoint $f\in\mathcal H_t$, let $\mathcal T_f$ contain the
forearm/hand--torso pairs with a valid worst candidate. The driving depth
and scheduled state are
\begin{align}
 D_f&=\max\bigl(\{0\}\cup\{[-c_{t+f,k,0}]_+:k\in\mathcal T_f\}\bigr), \\
 q_f&=\operatorname{clip}_{[0,1]}\frac{D_f-d_0}{d_1-d_0}, \\
 u^{(0)}&=(1-1/\tau_u)u_{t-1},\qquad
 u^{(1)}=\max(u^{(0)},q_0),\qquad
 u_t=\begin{cases}u^{(1)},&h_t=0,\\\max(u^{(1)},q_{h_t}),&h_t>0,\end{cases} \\
 m(u_t)&=1-u_t(1-U_s^{-1}),\qquad
 V_t=\frac{\pi V}{180\sqrt3 F}\bigl(1+u_t(U_v-1)\bigr), \\
 A_t&=\frac{\pi A}{180\sqrt3 F^2}\bigl(1+u_t(U_a-1)\bigr),\qquad
 c_{\rm cap}(u)=c\bigl(1+u(U_c-1)\bigr).
 \label{eq:urgency}
\end{align}
Missing candidates contribute no depth. The depth is measured after the
soft budget, fixed allowance, and clearance offset, rather than from the
raw signed gap. The rise follows the current depth immediately; the stored
state decays geometrically each non-idle internal step. The ramp is disabled
if $d_1\leq d_0$. The frozen schedule endpoints are
$d_0=\GuardUrgDZero$, $d_1=\GuardUrgDOne$, with
$\tau_u=\GuardUrgTau$.

The factor $m$ scales angular residual-velocity and residual-acceleration
costs and both Cartesian temporal costs. It leaves residual magnitude,
end-effector position/orientation, and pull weights unchanged.
Speed, acceleration, and the deep-row attenuation scale increase as shown.
The order of assembly matters: current penetration rows use
$c_{\rm cap}(u^{(1)})$; future rows, quadratic costs, and motion bounds use
the final $u_t$. Idle pass-through resets urgency to zero.

\subsection{Skin deformation: resolve local compression after host skinning}
\label{sec:method-skin}
Layer~2 takes the host's final skinned vertices/normals and static topology,
part memberships, seam masks, soft thickness, and undirected part pairs.
The collider set may be the same mesh or a separately posed mesh.
Static groups pool collider vertices into patches; hands use a finer group
set against other hands. The interface prepares feathered part memberships
and masks near shared seams. The C implementation chooses one contact
owner per pair, preferring hands, then arms, then the part with fewer
collider vertices. Part boxes, bounding spheres, and spatial indices prune
queries. Scratch arrays are reused, but previous poses or displacements do
not influence the next call.

For a patch centroid $q$, a nearest opposing vertex is found within the
local search reach $R_d=\GuardSkinSearch$ plus patch radius. Broad opposing
surfaces use a weighted plane; hands retain the exact nearest point and
normal. With $d_*$ the nearest distance and
$r_s=\GuardSurfaceSupport$, define
\begin{align}
 K(z)&=(1-z)^3(1+3z)\,\mathbf1_{0\leq z<1},\qquad
 z_j=\max\left(0,\frac{\|q-c_j\|^2-d_*^2}{r_s^2}\right), \\
 \bar c&=\frac{\sum_j K(z_j)c_j}{\sum_j K(z_j)},\quad
 \bar n=\frac{\sum_j K(z_j)n_j}{\|\sum_j K(z_j)n_j\|},\quad
 \bar s_B=\frac{\sum_j K(z_j)s_j}{\sum_j K(z_j)}.
 \label{eq:skin-plane}
\end{align}
Normal normalization uses a numerical floor. For every member $i$ of the
owner patch, let $\phi_i=(c_i-\bar c)^\top\bar n$ and
$f_i=[-n_i^\top\bar n]_+$. Only members with $f_i>0$ and
$\|c_i-\bar c\|\leq R_d$ contribute. For pair seam mask $\mu_{ik}$,
\begin{align}
 \lambda_i&=f_i^2\mu_{ik},\qquad
 w_i=\exp(-\phi_i/t_d)\lambda_i,\qquad t_d=\GuardContactTemperature, \\
 \bar q&=\frac{\sum_iw_ic_i}{\sum_iw_i},\quad
 \bar p=\frac{\sum_iw_i(c_i-\phi_i\bar n)}{\sum_iw_i},\quad
 \bar s_A=\frac{\sum_iw_is_i}{\sum_iw_i},\quad
 d_{\rm raw}=[-(\bar q-\bar p)^\top\bar n]_+, \\
 d&=\min\left\{\GuardDepthGain\,d_{\rm raw}
 S(d_{\rm raw}/\GuardSkinOnset)
 \min\left(1,\frac{\sum_i\lambda_i}{\GuardFacingOnset}\right),
 \rho(\bar s_A+\bar s_B)\right\}, \\
 d_A&=d\frac{\bar s_A}{\bar s_A+\bar s_B},\qquad
 d_B=d\frac{\bar s_B}{\bar s_A+\bar s_B}.
 \label{eq:skin-depth}
\end{align}
An event requires positive total thickness and
$d>\GuardSkinDepthCutoff$. Thus softer material receives a larger share
of the compression. This stage does not update the skeleton.

Each accepted event emits opposite displacement fields: $+\bar n$ around
$\bar q$ on owner $A$, and $-\bar n$ around $\bar p$ on opposing part $B$.
The broad width is $w_b=r_{\rm rim}+2\sqrt{\GuardWidthDepth\,d}$;
for a hand side with allocated depth $d_P$ the width is
$w_P=\GuardHandWidth+\GuardHandWidthGain\sqrt{\GuardHandWidthDepth\,d_P}$.
The interface default is $r_{\rm rim}=\GuardSkinRim$.
For event $e$ acting on part $P$ at center $p_e$, with direction $n_e$,
allocated depth $d_e$, and width $w_e$, vertex $i$ receives
\begin{align}
 a_{ei}&=d_e\,\pi_{iP}\mu_{ik}
 K(\|v_i-p_e\|^2/w_e^2),\qquad
 \beta_{ei}=(\GuardBlendScale\,a_{ei})^4, \\
 \widetilde d_i&=\frac{\sum_e\beta_{ei}a_{ei}n_e}
 {\max(\sum_e\beta_{ei},\varepsilon_w)},\qquad
 d_i=\widetilde d_i\min\left(1,
 \frac{\rho s_i}{\max(\|\widetilde d_i\|,\varepsilon_d)}\right).
 \label{eq:skin-field}
\end{align}
Here $\pi_{iP}$ is feathered part membership, and contributions
$a_{ei}<\GuardFieldCutoff$ are skipped. The numerical floors in the
denominator and norm match the implementation. The per-vertex cap is
applied after field blending.

A final local scaling step controls triangle folding. For original edges
$e,h$, normal $n=e\times h$, and displacement differences $z,w$, define
\begin{align}
 B_f&=\left[-\frac{(e\times w+z\times h)^\top n}
 {\max(\|n\|^2,\varepsilon_f)}\right]_+,\qquad
 C_f=\left[-\frac{(z\times w)^\top n}
 {\max(\|n\|^2,\varepsilon_f)}\right]_+, \\
 \eta_f&=\GuardAreaNumerator/2,\qquad
 s_f=\min\left(1,\frac{2\eta_f}
 {\max(\varepsilon_b,B_f+\sqrt{B_f^2+4\eta_f C_f})}\right).
 \label{eq:skin-fold}
\end{align}
For a nondegenerate triangle above the denominator floor, common scaling
by $s$ bounds its oriented-area ratio below by $1-B_fs-C_fs^2$.
Up to $\GuardAreaIterations$ passes scale moved vertices by their minimum
incident $s_f$ whenever a face scale is below $\GuardAreaTolerance$.
The code then recomputes a common minimum over affected faces and applies
it to all moved vertices if necessary. These reductions preserve the
per-vertex displacement cap; the final output is $v_i+d_i$ after scaling.
This local compression procedure does not guarantee removal of arbitrary
deep or global intersections.

The inter-layer budget is a static convention, not a transmitted remaining
slack. Layer~1 charges pose correction only after its part-dependent
$B_{ab}$ and fixed pair allowance are spent; Layer~2 re-detects contact and
caps each displacement by the global $\rho s_i$. In particular Layer~2
receives neither the smaller $\rho_h,\rho_f$ nor the pair allowances,
row slacks, or urgency. Even the Layer~1 no-penalty condition on a negative
gap permits
\begin{equation}
 -g_{ab}\leq B_{ab}+a_{t,k}-\epsilon
 \quad\text{when }a_{t,k}\geq\epsilon.
 \label{eq:budget-handoff}
\end{equation}
Consequently the hand-off reserves local compression for the skin stage,
while the code provides no exact budget equality or complete-resolution
guarantee between the stages.

\subsection{Cloth thickness: lift the target collision surface}
\label{sec:method-cloth}
An optional nonnegative rest cloth offset $o_i$ is stored at sampled body
vertices. A host-chosen $\kappa\in[0,1]$ modifies only arm--torso and
forearm/hand--leg pair classes. At their nearest target collider vertex,
\begin{align}
 \ell_b^{\rm cloth}&=(1-\kappa)o_b,\qquad
 g^{\rm cloth}_{ab}=g_{ab}-\ell_b^{\rm cloth},\qquad
 e^{\rm cloth}_{ab}=\begin{cases}
 \min(0,g^{\rm cloth}_{ab}+B_{ab}),&g^{\rm cloth}_{ab}<0,\\
 g^{\rm cloth}_{ab},&g^{\rm cloth}_{ab}\geq0,
 \end{cases} \\
 r_p^{\rm cloth}&=r_p+(1-\kappa)o_b,\qquad
 E_k=(1-\kappa)\max_{b:\,\mathsf p(b)=b_k,\ \text{collider}}o_b.
 \label{eq:cloth}
\end{align}
For these pairs, AABB reach, positive search cutoff, and adaptive dense
reach increase by $E_k$. The negative-plane inside test is unchanged;
row normals remain the original body-plane or point-separation normals.
Source observations always use zero lift. Contact pull uses the lifted
rest distance, including in its spatial gate.

In the negative lifted-gap branch, depth measured from the full cloth
surface, $o_b-g_{ab}$, may reach
$B_{ab}+\kappa o_b+a_{t,k}-\epsilon$ before a penetration hinge is
violated. Thus $\kappa$ specifies the compressible portion of the stored
offset. It is a collision-surface convention, not garment dynamics.
At $\kappa=1$ the lift vanishes and the body formulation is recovered.
Layer~2 continues to deform body skin with its original soft budget;
$\kappa$ is not passed to it.

\subsection{Parameters}
\label{sec:method-parameters}
Table~\ref{tab:method-parameters} gives the frozen configuration. All
reported values are generated from the \texttt{settings} object in the
release configuration; pack and embedded geometry constants are extracted
from their implementation by the same generator. Four unused layout fields
(\texttt{margin}, \texttt{calibration\_scale}, \texttt{source\_scale}, and
\texttt{positive\_offsets}) are zero and are not mechanisms in the method.
Lookahead and cloth compressibility are explicit host inputs rather than
entries in that frozen object.

\begin{table}[t]
 \centering\small
 \GuardParameterTable
 \caption{Frozen parameters. Temporal objective weights act on frame
 differences; speed and acceleration limits use physical time units before
 conversion. Angles in the clearance controls use degrees.}
 \label{tab:method-parameters}
\end{table}

\section{Evaluation}
\label{sec:eval}

\subsection{Data and protocol}

We evaluate the module on recorded output of a deployed motion generator
and on motion estimated from conversation video in the Seamless Interaction
dataset \cite{agrawal2025seamless}. The generator set holds \RgenClips{}
clips of \RgenFrames{} frames at 25\,fps. The video set holds 720 clips of
300 frames at 30\,fps. Bodies are SMPL-H shapes
\cite{loper2015smpl,romero2017mano}.

Settings were frozen before the final test. The primary result uses
\NClips{} clips, called unseen here, whose identifiers are disjoint from
every development, tuning, earlier evaluation, and urgency design set.
They run on three body shapes, giving \NClipBody{} clip--body pairs, with
each clip's own body as the source. The module and its settings were not
changed after these clips were first evaluated. We
also report sets from development, which shaped the design, and we mark
them as such.

\subsection{Measuring depth on the final skin}

Body depths are raw. They use no soft allowance and no source allowance,
and each one is measured on the mesh of the stage it reports, so Final is
the skin after Layer~2. The clothing experiment is the exception, and its
table says so.

The surface scan used throughout development takes a query vertex, finds
its nearest vertex on the other part, and reads depth along that vertex
normal. It is fast and it serves well on undeformed skin: among audited
copied-motion frames that it places above 20\,mm for hands, the solid test
below confirms \ScanConfirmCopyHand\%. On the final skin the assumption
fails. A contact dent bends the nearest normal, and a limb in a crease,
such as a hand on the chest with the arms crossed, can read as deep
inside a surface that it does not enter.

We therefore report an independent solid test. We close the torso mesh
by adding a centroid fan over each open boundary. A limb vertex is inside
when the generalized winding number of that closed mesh exceeds $0.5$,
and its depth is the exact point-to-triangle distance to the mesh. The
test uses no normals and no part of the runtime. We apply it to every
one of the \NFullFrames{} clip--body frames of the final skin. For Copy
and Pose we apply it to frames 7, 22, \ldots, 292 of each clip, which
gives \NUniformFrames{} clip--body frames per region and method, and to
the deepest frames that the scan flags in each clip. This test was added
after the first pass over the unseen clips, when scan maxima on the final
skin looked implausible and rendered frames showed no such depth. It
changes the measurement only. The scan places
\ScanFinalHandGtTwentyFrames{} final hand frames above 20\,mm, and the
solid test finds no final hand frame above \HandTorsoFinalMaxMM\,mm.

\subsection{Penetration}

Table~\ref{tab:depth} gives the result on unseen clips. Across all three
bodies, copied hands enter the torso beyond 20\,mm in
\HandTorsoCopyGtTwenty\% of frames, with a maximum of
\HandTorsoCopyMaxMM\,mm. The final skin reduces this to
\HandTorsoFinalGtTwenty\%, with a maximum of
\HandTorsoFinalMaxMM\,mm over every frame. For the forearm the rate falls from
\ForearmTorsoCopyGtTwenty\% to \ForearmTorsoFinalGtTwenty\%, with a
maximum of \ForearmTorsoFinalMaxMM\,mm.

The two layers do different work. Layer~1 removes deep conflicts: hand
frames beyond 20\,mm fall to \HandTorsoPoseGtTwenty\%, while hand frames
beyond 5\,mm remain at \HandTorsoPoseGtFive\%, which is the shallow
overlap left for tissue. Layer~2 then absorbs that overlap, and hand
frames beyond 5\,mm fall to \HandTorsoFinalGtFive\%.

The upper arm against the torso is the deliberate exception. Pose
correction leaves it close to the input, because forcing this contact
apart lifts the arm, and it is the region where real soft tissue yields.
The rate beyond 20\,mm is \UpperArmTorsoCopyGtTwenty\% for Copy,
\UpperArmTorsoPoseGtTwenty\% after Layer~1, and
\UpperArmTorsoFinalGtTwenty\% after Layer~2, with
\UpperArmTorsoFinalGtTen\% of final frames still beyond 10\,mm and a
maximum of \UpperArmTorsoFinalMaxMM\,mm.

Contacts outside the torso run through the surface scan. Hands against
other body parts fall from \HandOtherCopyGtTwenty\% to
\HandOtherFinalGtTwenty\% of frames beyond 20\,mm, and thigh against
thigh falls from \ThighCopyGtTwenty\% to \ThighFinalGtTwenty\%. Under the
solid test, \RegressionCount{} of \RegressionTotal{} clip--body--region
cases have a final maximum beyond 10\,mm that also exceeds the audited
copy maximum by more than 5\,mm.

\begin{table}[t]
\centering
\small
\begin{tabular}{@{}llrrrrrrr@{}}
\toprule
& & \multicolumn{3}{c}{Frames $>20$\,mm (\%)} & \multicolumn{2}{c}{Frames $>10$\,mm (\%)} & \multicolumn{2}{c}{Max (mm)} \\
\cmidrule(lr){3-5}\cmidrule(lr){6-7}\cmidrule(l){8-9}
Body & Region & Copy & Pose & Final & Copy & Final & Copy & Final \\
\midrule
Medium & Hand--torso & 9.47 & 0.09 & 0.00 & 12.88 & 0.00 & 115.0 & 5.4 \\
Medium & Forearm--torso & 5.19 & 0.00 & 0.00 & 9.80 & 0.00 & 62.2 & 5.3 \\
Medium & Upper arm--torso & 10.62 & 9.34 & 0.00 & 51.06 & 3.56 & 41.7 & 18.1 \\
\addlinespace
Tall heavy & Hand--torso & 16.75 & 0.34 & 0.00 & 21.78 & 0.00 & 133.9 & 6.3 \\
Tall heavy & Forearm--torso & 9.53 & 0.00 & 0.00 & 15.19 & 0.00 & 86.5 & 3.4 \\
Tall heavy & Upper arm--torso & 53.98 & 52.66 & 0.32 & 88.12 & 18.92 & 55.0 & 26.6 \\
\addlinespace
Short heavy & Hand--torso & 20.36 & 0.11 & 0.00 & 26.41 & 0.00 & 135.6 & 9.3 \\
Short heavy & Forearm--torso & 16.52 & 0.00 & 0.00 & 29.45 & 0.00 & 90.3 & 4.2 \\
Short heavy & Upper arm--torso & 54.77 & 51.86 & 0.02 & 87.44 & 8.24 & 58.5 & 27.1 \\
\bottomrule
\end{tabular}
\caption{Depth on the final skin over 320 unseen clips, by the independent solid test with no soft allowance. Copy is the input motion on the new body, Pose is Layer~1 alone, and Final is Layers 1 and 2. Layer~1 uses five frames of lookahead at 30\,fps. Copy and Pose rates use frames 7, 22, \ldots, 292 of each clip, and their maxima add the deepest frames flagged by the surface scan. Final rates and maxima use every frame.}
\label{tab:depth}
\end{table}

\begin{figure}[t]
 \centering
 \includegraphics[width=\linewidth]{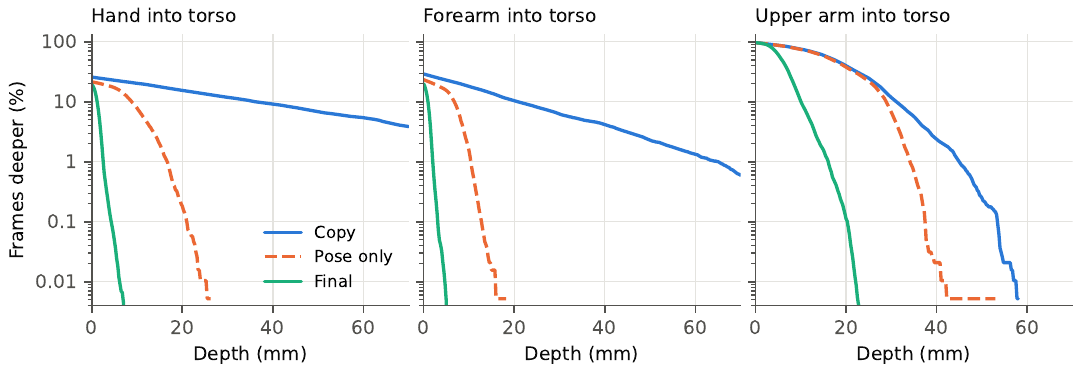}
 \caption{Share of frames deeper than a given depth on the unseen clips,
 pooled over the three bodies, under the solid test. Copy and Pose use
 the fixed-stride frames and Final uses every frame. Layer~1 removes deep
 conflicts and Layer~2 absorbs the shallow overlap that remains.}
 \label{fig:depth}
\end{figure}

\begin{figure}[t]
 \centering
 \includegraphics[width=\linewidth]{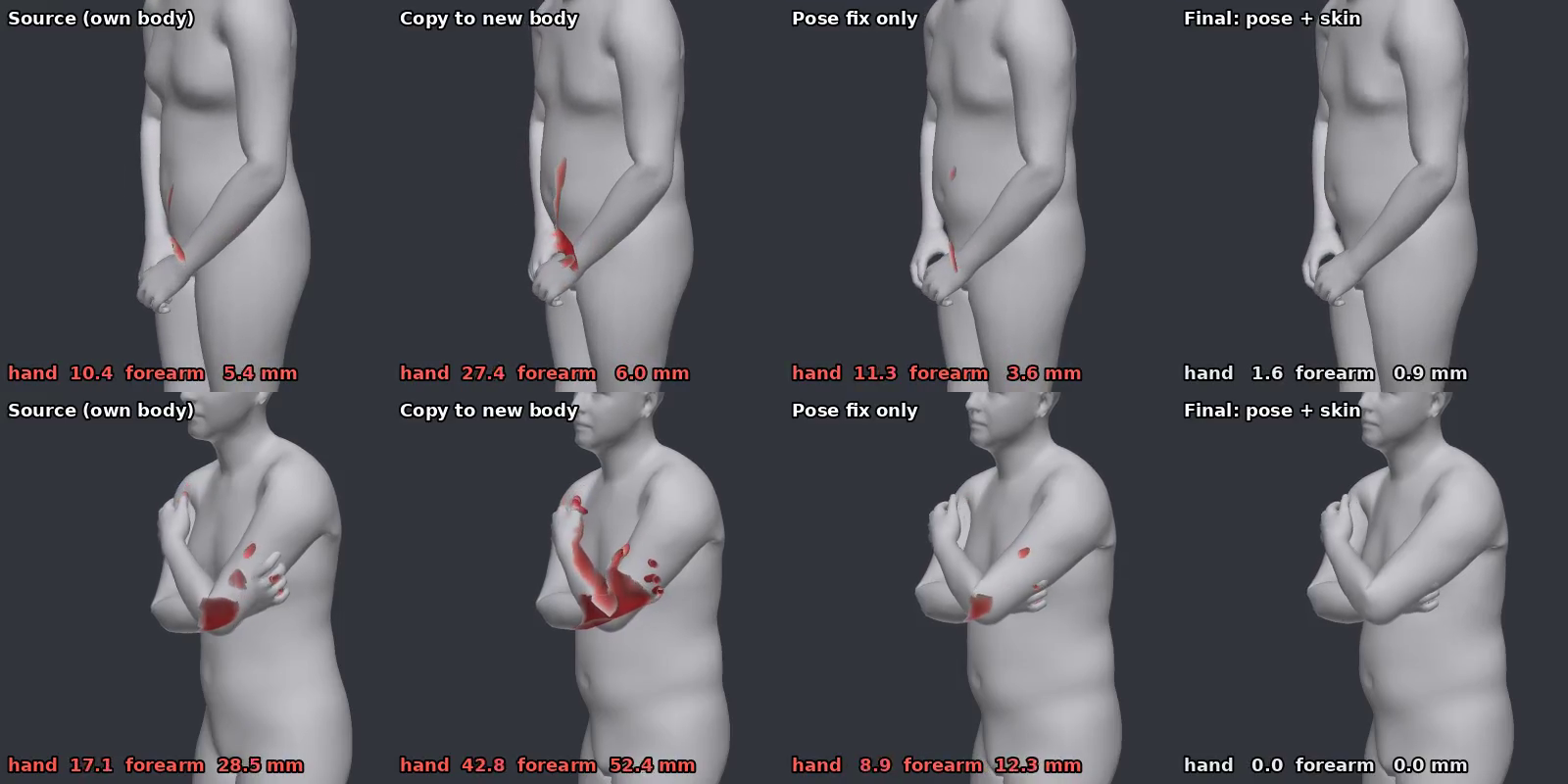}
 \caption{Two cases on the Tall heavy body. From the left: source on its
 own body, copied motion, Layer~1 alone, and the final skin. Top:
 generator output with a hand against the stomach. Bottom: crossed arms from
 an unseen clip, where the surface scan reports a deep hand on the final
 skin and the solid test finds none.}
 \label{fig:cases}
\end{figure}

\subsection{Motion preservation}

Table~\ref{tab:natural} gives the cost of correction. The median
high-frequency ratio over all pairs is \HFMedian{} with a 90th percentile
of \HFPNinety{}, so correction adds little jitter; the worst single pair
reaches \HFMax{}. The mean wrist displacement from copied motion is
\HandOffsetMeanCM\,cm. Source-labelled contact pair--frames satisfy the
pose-stage contact criterion in a share of \ContactKeptFinal{}, against
\ContactKeptCopy{} for copied motion. \UntouchedPct\% of joint--frame
rotations stay within $0.5^\circ$ of the input, and \ExactFramePct\% of
frames carry an exactly zero correction on every joint. These clips are dense in hand--body
contact, so that share reflects the data as well as the module.

\begin{table}[t]
\centering
\small
\begin{tabular}{@{}lrrrrrrrr@{}}
\toprule
& \multicolumn{2}{c}{Wrist offset (cm)} & \multicolumn{2}{c}{HF ratio} & Arm & \multicolumn{2}{c}{Retained} & Joints \\
\cmidrule(lr){2-3}\cmidrule(lr){4-5}\cmidrule(lr){7-8}
Body & Mean & P95$^{\dagger}$ & Median & P90 & speed & Pull & Contact & ${<}0.5^\circ$ (\%) \\
\midrule
Medium & 0.68 & 1.54 & 1.008 & 1.058 & 1.008 & 0.962 & 0.954 & 83.1 \\
Tall heavy & 0.76 & 1.73 & 1.007 & 1.063 & 1.007 & 0.948 & 0.943 & 83.5 \\
Short heavy & 1.10 & 2.37 & 1.010 & 1.095 & 1.009 & 0.953 & 0.961 & 77.9 \\
\bottomrule
\end{tabular}
\caption{Motion preservation on the unseen clips with five frames of lookahead at 30\,fps. The high-frequency (HF) ratio is the per-clip mean, over ten arm and leg joints, of the corrected-to-copied ratio of joint-velocity energy at or above 3\,Hz, so 1 means equal measured energy. Retained pulls and contacts are weighted over active pair--frames. The last column is the share of joint--frame rotations within $0.5^\circ$ of the input. $^{\dagger}$Averaged over per-clip P95 values.}
\label{tab:natural}
\end{table}

\subsection{Deployed generator output}

Table~\ref{tab:model} replays recorded output of a deployed generator
through the module, which is its intended position in a pipeline. The
final surface scan records no hand--torso frame above 20\,mm across the
five bodies. The Extreme heavy body uses three times
the shape coefficients of Short heavy, and it stays free of deep
hand--torso frames at the cost of a much larger hand offset.

\begin{table}[t]
\centering
\small
\begin{tabular}{@{}lrrrrrrrr@{}}
\toprule
& \multicolumn{2}{c}{Hand--torso $>20$\,mm (\%)} & \multicolumn{2}{c}{Hand--torso max (mm)} & \multicolumn{2}{c}{Forearm--torso max (mm)} & Offset & HF \\
\cmidrule(lr){2-3}\cmidrule(lr){4-5}\cmidrule(lr){6-7}
Body & Copy & Final & Copy & Final & Copy & Final & (cm) & median \\
\midrule
Medium & 0.76 & 0.00 & 106.8 & 4.0 & 32.9 & 3.7 & 0.15 & 1.002 \\
Tall slim & 0.22 & 0.00 & 87.9 & 3.8 & 14.5 & 3.3 & 0.15 & 1.001 \\
Tall heavy & 5.26 & 0.00 & 127.1 & 4.2 & 34.1 & 5.1 & 0.16 & 1.001 \\
Short heavy & 7.10 & 0.00 & 138.2 & 4.3 & 51.4 & 3.0 & 0.36 & 1.003 \\
Extreme heavy & 35.56 & 0.00 & 173.9 & 5.0 & 95.2 & 4.4 & 3.51 & 1.043 \\
\bottomrule
\end{tabular}
\caption{Recorded output from a deployed motion generator: 64 clips of 752 frames at 25\,fps, replayed with four frames of lookahead. Depths use the surface scan on the final skin.}
\label{tab:model}
\end{table}

\subsection{Harder sets from development}

Two development sets show where the module pays more. Layer~1 alone
takes the Extreme heavy body from \ExtremeCopyGtTwenty\% to
\ExtremePoseGtTwenty\% of hand frames beyond 20\,mm on \ExtremeClips{}
clips, and this body has little room left for the motion: the mean hand
offset grows to \ExtremeHandOffsetCM\,cm, and intended hand pulls are
retained at \ExtremePullKept{}. On AMASS motion \cite{mahmood2019amass} on
an SMPL-X \cite{pavlakos2019smplx} Tall heavy body, Layer~1 takes hand
frames beyond 20\,mm from \AmassRandomCopyGtTwenty\% to
\AmassRandomPoseGtTwenty\% on \AmassRandomClips{} random clips, and from
\AmassHardCopyGtTwenty\% to \AmassHardPoseGtTwenty\% on \AmassHardClips{}
hard clips. These figures use the surface scan.

\subsection{Depth-triggered urgency}

Table~\ref{tab:urgency} isolates the urgency schedule. Strong temporal
smoothing removes oscillation on resting contact, but on its own it
reacts too slowly to a fast entry. Hand and forearm clearance against
the torso drives a shared urgency state, which adjusts temporal costs and
motion bounds for the whole pose and can persist after the triggering
conflict. The maximum depth falls sharply. The 90th percentile of the
high-frequency ratio rises slightly, which is the cost of responding
faster.

\begin{table}[t]
\centering
\small
\begin{tabular}{@{}lrrrrrr@{}}
\toprule
& \multicolumn{2}{c}{Hand--torso $>20$\,mm (\%)} & \multicolumn{2}{c}{Hand--torso max (mm)} & \multicolumn{2}{c}{HF ratio P90} \\
\cmidrule(lr){2-3}\cmidrule(lr){4-5}\cmidrule(l){6-7}
Clips & Without & With & Without & With & Without & With \\
\midrule
Medium & 0.25 & 0.05 & 57.9 & 30.3 & 1.041 & 1.047 \\
Tall heavy & 0.24 & 0.02 & 87.7 & 23.9 & 1.029 & 1.061 \\
Short heavy & 0.29 & 0.03 & 81.0 & 29.1 & 1.047 & 1.081 \\
Tall heavy, hard & 0.31 & 0.03 & 64.6 & 23.5 & 1.094 & 1.113 \\
\bottomrule
\end{tabular}
\caption{Layer~1 alone with and without the urgency schedule, by surface scan, on the set that exposed fast entries. Layer~1 uses five frames of lookahead at 30\,fps. Each random row has 110 clips and the hard row has 80. These clips informed the schedule design, so they are a design record, not a held-out result.}
\label{tab:urgency}
\end{table}

\subsection{Clothing}

Table~\ref{tab:cloth} runs six generated bodies with fitted garments.
Lowering $\kappa$ moves the arm off the garment surface instead of
letting it press the cloth flat. These clips come from the development
set, and the table reports Layer~1 alone with the allowances that its
caption states.

\begin{table}[t]
\centering
\small
\begin{tabular}{@{}lrrrrrr@{}}
\toprule
& \multicolumn{4}{c}{Frames with adjusted cloth excess $>5$\,mm (\%)} & \multicolumn{2}{c}{$\kappa=0$} \\
\cmidrule(lr){2-5}\cmidrule(l){6-7}
Body & Copy & $\kappa=1$ & $\kappa=0.5$ & $\kappa=0$ & Offset (cm) & HF \\
\midrule
Heavy A & 30.00 & 19.63 & 6.43 & 0.03 & 0.87 & 1.025 \\
Slim & 0.00 & 0.00 & 0.00 & 0.00 & 0.29 & 1.007 \\
Muscular & 1.57 & 1.17 & 0.23 & 0.00 & 0.27 & 1.007 \\
Stocky & 1.53 & 0.00 & 0.00 & 0.00 & 0.45 & 1.009 \\
Tall thin & 0.50 & 0.33 & 0.33 & 0.30 & 0.30 & 1.009 \\
Heavy B (dress) & 2.87 & 0.00 & 0.00 & 0.00 & 0.74 & 1.019 \\
\bottomrule
\end{tabular}
\caption{Six generated bodies with fitted garments on ten development clips, after Layer~1 alone, with five frames of lookahead at 30\,fps. Excess is measured after soft and source allowances, unlike the raw depths elsewhere. Smaller $\kappa$ treats the garment as harder.}
\label{tab:cloth}
\end{table}

\subsection{Latency}

Times are measured on one pinned core of an AMD EPYC 7K62, with other
jobs running on the machine. Excluding the first frame of each clip, with
five frames of lookahead, mean Layer~1 cost averages \LayerOneMeanMS\,ms
across the SMPL-H body groups. The largest group 99th percentile is
\LayerOnePNinetyNineMS\,ms and the maximum is \LayerOneMaxMS\,ms. One frame
of the SMPL-X run took \LayerOneOutlierMS\,ms. The causal setting averages
\LayerOneCausalMeanMS\,ms. Layer~2 averages \LayerTwoMeanMS\,ms on a mesh of
\LayerTwoVertices{} vertices, with a largest group 99th percentile of
\LayerTwoPNinetyNineMS\,ms. The sum of these separately measured mean stage
costs is \TotalMeanMS\,ms, which leaves most of a 33\,ms frame to the host.
It excludes host skinning and interface overhead. Lookahead adds $L/F$ seconds of buffering
delay, independent of processing time.

\subsection{Reproduction}

\begin{figure}[t]
 \centering
 \includegraphics[width=\linewidth]{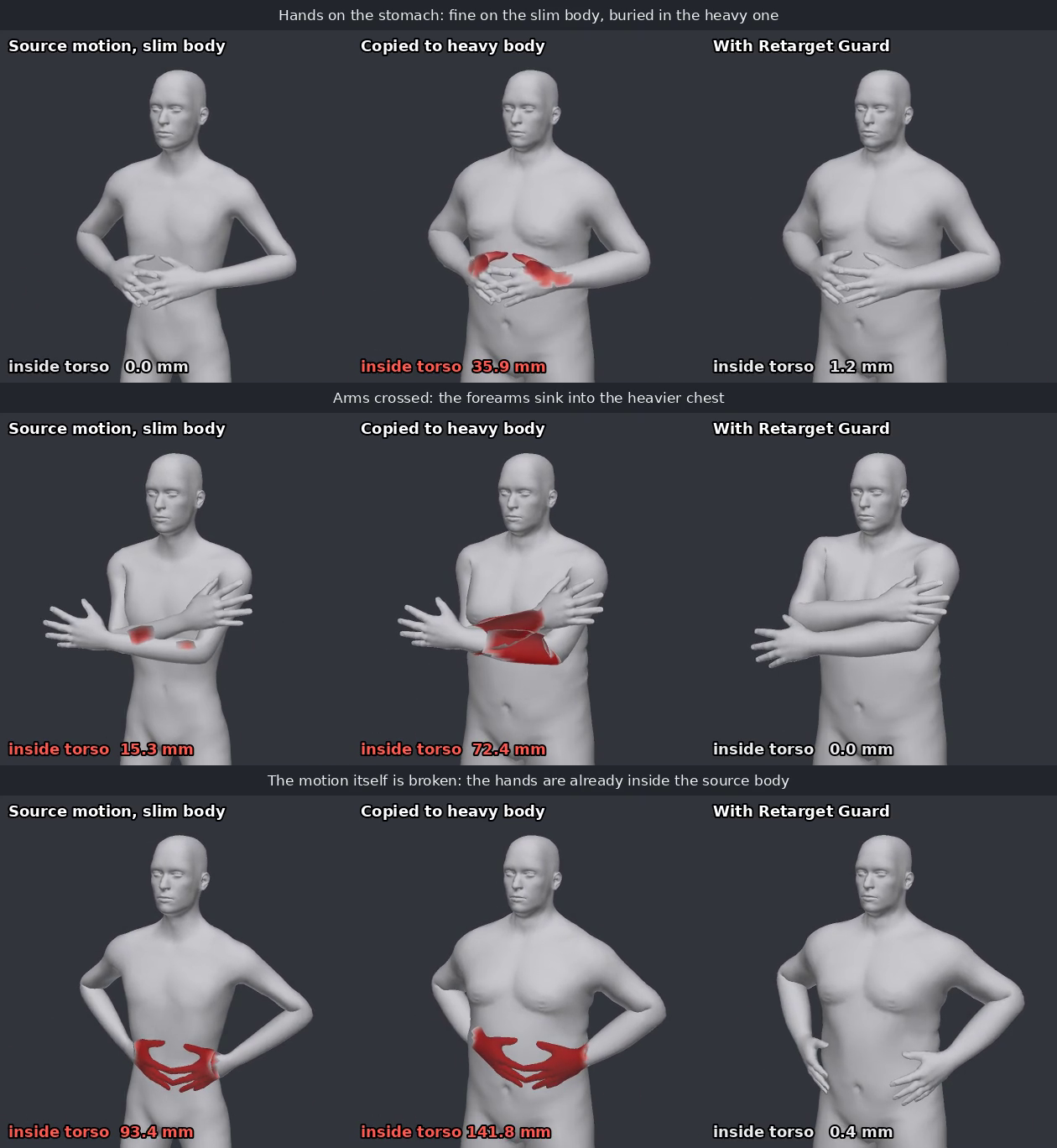}
 \caption{Public demonstration on CC0 bodies. Red marks the part of the
 hand and forearm inside the torso. Labels give the deepest vertex under
 the solid test. Top: hands on the stomach. Middle: crossed arms. Bottom:
 a source motion that already penetrates its own body.}
 \label{fig:demo}
\end{figure}

The module, the build tool, the tests, and a demonstration that needs no
licensed data are released with this report at
\url{https://github.com/zhuyifeiabcd1/retarget-guard}. The demonstration uses
CC0 characters from MPFB \cite{mpfb2026} and procedural motion.
Figure~\ref{fig:demo} shows it.

\section{Discussion}
\label{sec:discussion}

The module targets local self-penetration repair after motion generation
or retargeting. Its separation into pose correction and skin compression
lets the host preserve its generator and rendering choices while
allocating shallow contact to deformable tissue. This interface requires
a native character pack for the motion model and a target pack for each
skin. Both packs must provide compatible semantic joints and collision
pairs; native poses require the appropriate skeleton and bind conventions.
Model independence does not remove this per-character preparation.
Layer~1 uses sampled geometry with truncated skin influences, so a host
with different skinning or additional shape deformation may produce a
different surface. Layer~2 observes the supplied final vertices directly.

The correction is local and bounded. Sampled nearest-vertex normal planes
are collision surrogates, and the prescribed part pairs do not cover
every possible intersection. Penetration and pull are soft costs, the
solver uses fixed work, and release and snapping follow the solve.
There is no formal feasibility certificate or guarantee of removing
arbitrary deep or global intersections. Root translation, absolute
anatomical joint limits, balance, and environmental contact are outside
the correction problem. Uncontrolled local rotations, including fingers,
are copied, although their world transforms inherit corrected ancestors.

The static budget convention expresses an allocation of correction
between the layers. It does not transmit remaining slack: the pose layer
uses part-dependent fractions and fixed pair allowances, whereas the
skin layer uses the global fraction and independently detects contact.
Skin compression is geometric, without friction, elastic dynamics, or
volume conservation. Cloth thickness/compliance modifies selected pose
collision tests; it neither simulates garments nor changes Layer~2's
body-skin budget. Short lookahead adds buffering delay and checks target
endpoints rather than an entire intervening trajectory. Frame-exact
pass-through applies to clean, inactive zero-state frames at the native
interface; earlier correction state returns through release.

The tests use parametric bodies and generated garments. Field garments
have not been evaluated. Windows and macOS have not been verified, and
the module has not yet been integrated with a shipping motion model.
The current evaluation therefore addresses the isolated repair module
under its stated host interface. Future work includes production
integration, cross-platform testing, and evaluation on captured garments
and artist-authored skins. A measured budget exchange between layers,
richer host-surface derivatives, and adaptive collision coverage could
improve difficult contacts. Explicit trajectory checks would extend the
temporal scope beyond the present endpoint surrogate.


\end{document}